\documentstyle[aps,prl,epsfig,multicol]{revtex} 

\newcommand{\GPE}{Gross-Pitaevskii equation } 
\begin{document}

\title{Simulations of Bose fields at finite temperature}
\author{M.~J.~Davis, S.~A.~Morgan, and K.~Burnett}
\address{Clarendon Laboratory, Department of Physics, University of Oxford,
Oxford OX1~3PU, United Kingdom}
\maketitle

\begin{abstract} 

We introduce a time-dependent projected Gross-Pitaevskii equation to describe a
partially condensed homogeneous Bose gas, and find that this equation will
evolve randomised initial wave functions to equilibrium.  We compare our
numerical data to the predictions of a gapless, second order theory of
Bose-Einstein condensation [S.~A.~Morgan, J.~Phys.~B {\bf 33}, 3847 (2000)],
and find that we can determine a temperature when the theory is valid.   As
the Gross-Pitaevskii equation is non-perturbative, we expect that it can describe
the correct thermal behaviour of a Bose gas as long as all relevant modes are
highly occupied.  Our method could be applied to other boson fields.

\end{abstract}

\begin{multicols}{2}

The achievement of Bose-Einstein condensation (BEC) in a dilute gas offers the
possibility of studying the dynamics of a quantum field at finite temperatures
in the laboratory \cite{JILA,MIT}.  However, direct numerical simulation of
the full equations of motion for such systems is well beyond the capability of
today's computers.  Even equilibrium calculations in the region of a phase
transition require non-perturbative methods, meaning that fully quantal
treatments are unfeasible.  

At finite temperature, when there are an appreciable number of non-condensed
particles, the fully quantal second order  theory of Morgan \cite{sam}  (and
other equivalent treatments \cite{fedichev,giorgini})  should be sufficient for
an accurate description of many properties of the dilute Bose gas in
equilibrium and away from the region of critical fluctuations.   Dynamical
treatments are much harder, and in general require significant approximations.
For example, calculations have been performed for small systems 
\cite{Drummond}; with a restricted number of modes \cite{holland}; and for the
dynamics of condensate formation where the ground state is assumed to grow
adiabatically \cite{Gardiner}.

The Gross-Pitaevskii equation (GPE) has been used to predict the properties of
condensates near $T=0$, when there are very few non-condensate atoms present. 
Both statically and dynamically it has shown excellent agreement with experiment
\cite{Edw95,Rup95,dalfovo}. 
It has been argued, however, that the GPE can be used to describe the dynamics
of a Bose-Einstein condensate at finite temperature
\cite{boris,kagan1,Marshall}.  In the limit where the modes of the system are
highly occupied ($N_k \gg 1$), the classical fluctuations of the field
overwhelm the quantum fluctuations, and these modes may therefore be
represented by a coherent wave function.  This is analogous to the situation in
laser physics, where the highly occupied laser modes can be well described by
classical equations.  

Using this argument,  Damle {\em et al.}\ have performed
calculations of the approach to equilibrium of a near ideal superfluid
\cite{damle}, and similar approximations to other quantum field equations have
been successful elsewhere \cite{Turok}.  References 
\cite{bijlsma,goral,sinatra} also use the GPE to represent the classical modes
of a Bose-condensed system. The main advantage of this method is that realistic
calculations, while still a major computational issue, are feasible---methods
for solving the GPE are well developed.  Also, as the GPE is
non-perturbative it should be possible to study the region of the phase
transition, as long as the condition $N_K \gg 1$ is satisfied. 

There are, however, problems associated with the GPE.  It is a classical
equation, and so in equilibrium it will satisfy the equipartition theorem---all
modes of the system will contain an energy $k_B T$.  Thus, if we couple a
system to a heat bath, and solve the equation with infinite
accuracy, we will observe an ultra-violet catastrophe.  Also, the higher the
energy of any given mode, the lower its occupation will be in equilibrium, and
eventually the criterion  $N_k \gg 1$ will no longer be satisfied.  For these
low occupation modes a form of kinetic equation is more appropriate.  The
solution to both of these problems is to introduce a {\em cutoff} in the modes
represented by the GPE.

%For the GPE to be accepted for use at finite temperature, there needs to be
%quantitative evidence that it can  reproduce the results of existing quantum
%field theories for the Bose gas, and in particular, that a measure of
%temperature can be established.  In this letter we present the results of
%calculations that provide such evidence for a homogeneous gas.  We have derived
%a projected GPE for the highly occupied modes of the system and carried out
%simulations for a range of initial energies for a fixed number of atoms.  We
%observe the system to evolve to equilibrium, and we compare its properties to
%those predicted by the  finite temperature  theory of Ref.~\cite{sam}.

Our theoretical approach begins with the full operator equation for the Bose
field with two-body interactions
\begin{equation}
i \hbar\frac{\partial \hat{\Psi}({\bf r})}{\partial t} =
\hat{H}_0 \hat{\Psi}({\bf r}) + 
U_0
\hat{\Psi}^{\dag}({\bf r})
\hat{\Psi}({\bf r})\hat{\Psi}({\bf r}),
\label{eqn:field_operator}
\end{equation}
where $U_0 = 4\pi \hbar^2 a / m$ is the effective interaction
strength at low momenta,  $a$ is the {\em s}-wave scattering length, and $m$ is
the particle mass.  The route to the usual GPE is to assume that the full field
operator can be replaced by a wave function $\psi({\bf r})$---i.e.\ that {\em
all} quantum fluctuations can be neglected.  We proceed instead by defining a
projection operator $\hat{\mathcal{P}}$ such that
\begin{eqnarray}
\hat{\mathcal{P}}\hat{\Psi}({\bf r})  = 
\sum_{{\bf k} \in C} \hat{a}_{\bf k} \phi_{\bf k}({\bf r}),
\end{eqnarray}
where the region $C$ is {\em determined} by the 
requirement that $\langle \hat{a}_{\bf k}^{\dag} \hat{a}_{\bf k}\rangle \gg 1$,
and the set $\{\phi_{\bf k}\}$ defines some basis in which the field operator is
approximately diagonal at the boundary of $C$.
For these modes, the quantum fluctuation part of the projected field operator can be
ignored, and so we replace $\hat{a}_{\bf k} \rightarrow c_{\bf k}$ and write
\begin{eqnarray}
\psi({\bf r})  = \sum_{{\bf k} \in C} c_{\bf k} \phi_{\bf k}({\bf r}).
\end{eqnarray}

Defining the operator $\hat{\mathcal{Q}} = \hat{\openone} - \hat{\mathcal{P}}$ and 
$\hat{\mathcal{Q}}\hat{\Psi}({\bf r}) = \hat{\eta}({\bf r})$, operating on
Eq.~(\ref{eqn:field_operator}) with $\hat{\mathcal{P}}$ and taking the mean
value results in what we call the finite temperature GPE
\begin{eqnarray}
i \hbar \frac{\partial {\psi}({\bf r})}{\partial t}
&=& 
\hat{H}_0 \psi({\bf r}) + U_0
\hat{\mathcal{P}}\left\{|\psi({\bf r})|^2 \psi({\bf r})\right\} 
\nonumber
\\
&+&
U_0\hat{\mathcal{P}}\left\{2|\psi({\bf r})|^2 \langle \hat{\eta}({\bf r})\rangle
+
\psi({\bf r})^2\langle\hat{\eta}^{\dag}({\bf r})\rangle\right\}
\nonumber
\\
&+&
U_0\hat{\mathcal{P}}\left\{\psi^*({\bf r})
\langle\hat{\eta}({\bf r})\hat{\eta}({\bf r})\rangle
+
2\psi({\bf r})\langle\hat{\eta}^{\dag}({\bf r})\hat{\eta}({\bf r})\rangle\right\}
\nonumber
\\
&+&
U_0\hat{\mathcal{P}}\left\{\langle\hat{\eta}^{\dag}({\bf r})
\hat{\eta}({\bf r})\hat{\eta}({\bf r})\rangle
\right\}.
\label{eqn:ftgpe}
\end{eqnarray}
This describes the full dynamics of the region $C$ and its coupling to an
effective heat bath $\hat{\eta}({\bf r})$, which in principle 
can be described using a form of quantum kinetic theory.
The finite temperature GPE is discussed in detail in Ref.~\cite{formalism}.

In this letter, however, we wish to show that the GPE {\em alone} can describe the
evolution of general configurations of the coherent region $C$ towards an
equilibrium that can be parameterised by a temperature.  We therefore ignore
all terms involving  $\hat{\eta}({\bf r})$ in Eq.~(\ref{eqn:ftgpe}) and
concentrate on the first line, which we call the projected GPE.  Although this
equation is both unitary and reversible, we expect it to evolve general states
to equilibrium, because deterministic nonlinear systems exhibit
chaotic, and hence ergodic, behaviour if more than a few degrees of freedom are
present \cite{reichl}. This is confirmed by our numerical simulations and forms
the main result of this letter.

The projected GPE describes a microcanonical system.  However, if the region
$C$ is large, then fluctuations in energy and particle number in the  grand
canonical ensemble would be small.  Hence we expect the final  equilibrium
state of the projected GPE to be similar to that of the finite temperature GPE
coupled to a bath $\hat{\eta}({\bf r})$ with the appropriate chemical potential
and temperature. This does not affect the main result of this letter, however,
which is simply that equilibrium is attained. The detailed non-equilibrium
dynamics of the system {\em will} depend on the exchange of energy and
particles between $C$ and the bath, and this will be addressed in future work.

We have performed simulations for a fully three-dimensional homogeneous Bose
gas with periodic boundary conditions.  This choice has been made to simplify
the projection operation that must be carried out.  In this case the single
particle states are plane waves, and the effect of a condensate is simply to
mix modes of momenta ${\bf p}$ and $-{\bf p}$. This allows us to apply the
projector cleanly in momentum space, which is easily accessible by 
fast fourier transform.   In principle there is no barrier to performing the
same computation in a trap---in practice the projection operation is much more
time consuming.
%, and we hope to address this issue in future work.

The dimensionless equation we compute is
\begin{equation}
i \frac{\partial \psi(\tilde{ \bf r})}{\partial \tau}
= -\tilde{\nabla}^2 \psi( \tilde{\bf r}) + 
C_{\rm nl} \hat{\mathcal{P}}|\psi(\tilde{ \bf r})|^2\psi(\tilde{ \bf r}), 
\label{eqn:gpe}
\end{equation}
where we have defined 
$\int d^3\tilde{\bf r} |\psi(\tilde{\bf r})|^2 = 1$.  
The nonlinear constant is $C_{\rm nl} = {2 m N U_0}/{\hbar^2 L}$, where $N$ is the total
number of particles in the volume, and $L$ is the
period of the system.  Our dimensionless parameters are $\tilde{\bf r} = {\bf r}/L$, wave vector 
$\tilde{\bf k} = {\bf k} L$, energy 
$\tilde{\varepsilon} = \varepsilon / \varepsilon_L$, and 
time $\tau = \varepsilon_L t/ \hbar$, with 
$\varepsilon_{L} = \hbar^2/(2 m L^2)$.

The calculations presented here have been performed with $C_{\rm nl}=2000$, and
the projector $\hat{\mathcal{P}}$ chosen such that all modes have $|{\bf k}| < 15 \times 2\pi/L$. 
This means that most of the states contained in the calculation are phonon-like
for large condensate fraction.  We  note that while the number of states in the
problem is fixed, the nonlinear constant  only determines the ratio of $N U_0 /
L$.  This means that for a given value of $C_{\rm nl}$, we are free to choose $N$, $U_0$
and $L$ such that our condition $|c_{\bf k}|^2 \gg 1$ is always satisfied for a
given physical situation. In particular we can choose $^{87}$Rb atoms with
$N= 5 \times 10^5$ and $L\approx 17$ $\mu$m to give a number density of about
$10^{14}$ cm$^{-3}$---similar parameters to current experiments in traps.

\begin{figure}\centering
\epsfig{file=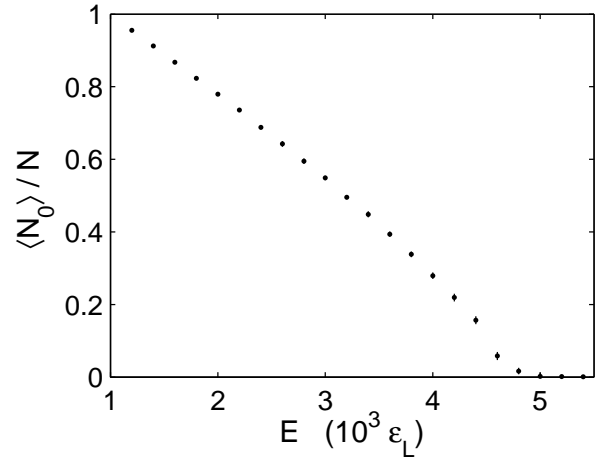,width=8cm}
\caption{Condensate fraction plotted against total energy after each
individual simulation has reached equilibrium.  The barely discernable
vertical lines on each point indicate the magnitude of the fluctuations.} 
\label{fig:n0ve}
\end{figure}

We begin our simulations with  wave functions far from equilibrium with a
chosen  total energy $\tilde{E}$.  They have a flat distribution in $k$-space
out to some maximum momentum determined by $\tilde{E}$, and the phase of each
momentum component is chosen at random.  These initial states are then
evolved for a time period of $\tau = 0.4$, by which stage equilibrium appears
to have been reached. We determine the properties of the system at equilibrium
by assuming that the ergodic theorem applies, and time-averaging over 50 wave
functions from the last $\delta \tau = 0.1$ of the simulation. We find that
the equilibrium properties depend only on the total energy---they are
independent of the details of the initial wave function.
 
\begin{figure}\centering
\epsfig{file=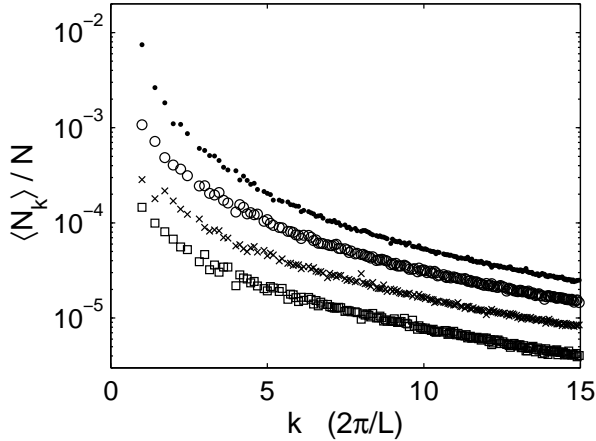,width=8cm}
\caption{Plots of the equilibrium Bogoliubov quasiparticle 
distributions averaged over time and angle for four different total energies. 
Squares, $\tilde{E} = 1600$; crosses, $\tilde{E} = 2000$; 
circles, $\tilde{E} = 3200$; dots, $\tilde{E} = 4600$.
The mean condensate occupation 
for all four distributions is off-axis.} 
\label{fig:qp_distribution}
\end{figure}

Strong evidence that the simulations have reached equilibrium is given by the
time dependence of the condensate population.  For all energies this settles
down to an average value that fluctuates by a small amount, and the results are
presented in Fig.~\ref{fig:n0ve}.  Further support is provided by the
distribution of the particles in momentum space.  Rather than using the
plane-wave basis, we transform the wave functions into the  quasiparticle basis
of quadratic Bogoliubov theory, the sole parameters of the transformation being
the condensate fraction and the nonlinear constant  $C_{\rm nl}$.  We then
average the populations of the quasiparticles states over time and angle to
produce a one dimensional plot, and the results are shown in
Fig.~\ref{fig:qp_distribution}.  

The GPE is the high occupation limit of the full equation for the Bose field
operator.  Therefore, in equilibrium we expect the mean occupation of the
quasiparticle mode ${ k}$ to be the classical limit of the Bose-Einstein 
distribution---i.e.\ the equipartition relation 
\begin{equation} \langle N_{k}
\rangle = \frac{k_B T}{\varepsilon_{ k} - \mu}. 
\label{eqn:occupation}
\end{equation} Since we can determine the Bogoliubov occupation 
$\langle N_{k} \rangle$ from our simulation data,
we can attempt to fit this distribution to a dispersion relation for
$\varepsilon_{ k}$, and hence determine the temperature.

\begin{figure}\centering
\epsfig{file=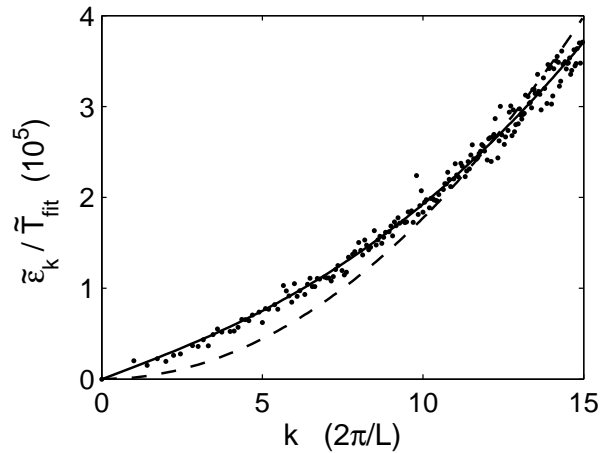,width=8cm}
\caption{
Comparison of dispersion relations for $\tilde{E} = 1400$ with 
$\langle N_0\rangle/N = 0.912$.  The dots are a plot of the RHS of
Eq.~(\protect\ref{eqn:fit}), and the lines plot $\tilde{\varepsilon}_k /
\tilde{T}_{\rm fit}$.  The solid line is for the 
Bogoliubov dispersion relation with 
$\tilde{T}_{\rm fit} = 0.0284$, while the dashed line is for the ideal gas with 
$\tilde{T}_{\rm fit} = 0.0223$.
} 
\label{fig:bog_dispersion}
\end{figure}

In the limit of large condensate fraction $\langle N_0 \rangle/N \sim 1$, we
expect the Bogoliubov dispersion relation to be a good estimate of the
energies.  The Bogoliubov transformation approximates the many-body Hamiltonian
by a quadratic form, which can be diagonalised exactly. The eigenstates are
quasiparticles, and in our dimensionless units the dispersion relation takes the
form
\begin{equation}
\tilde{\varepsilon}_{k} = \left(\tilde{k}^4 + 2 
C_{\rm nl}\frac{\langle N_0 \rangle}{N} \tilde{k}^2\right)^{1/2}.
\label{eqn:bog}
\end{equation}
Manipulating Eq.~(\ref{eqn:occupation}) and measuring the excitation spectrum
relative to the condensate we find
\begin{equation} 
\frac{\tilde{\varepsilon}_{k}}{\tilde{T}} = \left(\frac{N}{\langle N_{ k} \rangle} -
\frac{N}{\langle N_0 \rangle}\right), 
\label{eqn:fit} 
\end{equation} 
where $\tilde{T} =  k_B T/ (N \varepsilon_L)$  is our dimensionless
temperature, and the second term on the RHS arises from the difference between the
condensate energy and the chemical potential of the system.  By comparing the
curve of this relation with that of  Eq.~(\ref{eqn:bog}), a temperature can be
determined.  For the $\tilde{E} = 1400$ simulation we find $\tilde{T} = 0.0284$
gives an excellent fit, and this is shown in Fig.~\ref{fig:bog_dispersion}.  At
higher simulation energies, however, the shape of Eq.~(\ref{eqn:fit}) no
longer agrees with Eq.~(\ref{eqn:bog}) and we must use a more sophisticated
theory to predict the dispersion relation.

\begin{figure}\centering
\epsfig{file=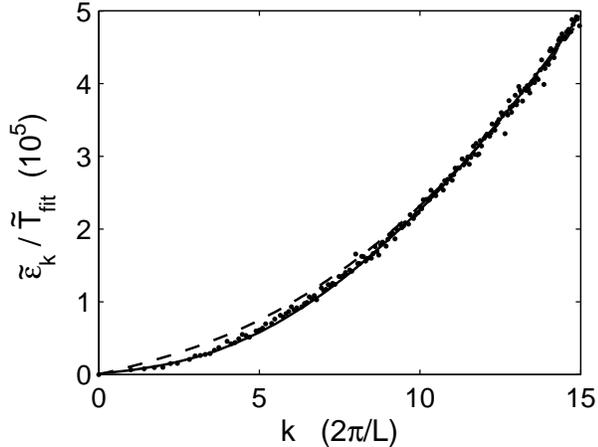,width=8cm}
\caption{
Comparison of dispersion relations for $\tilde{E} = 4000$ with 
$\langle N_0\rangle/N = 0.279$.  The dots are a plot of the RHS of
Eq.~(\protect\ref{eqn:fit}) and the lines plot $\tilde{\varepsilon}_k /
\tilde{T}_{\rm fit}$.  
The dashed line is for the Bogoliubov dispersion relation
with $\tilde{T}_{\rm fit} = 0.193$, and the solid line is
for the second order theory of Ref.~\protect\cite{sam} with 
$\tilde{T}_{\rm fit} = 0.201$.  
} 
\label{fig:full_theory}
\end{figure}

As the occupation of the quasiparticle modes becomes significant (in this case
more than a few percent), the cubic and quartic terms of the many-body
Hamiltonian that were neglected in the Bogoliubov transformation become
important.
In Ref.~\cite{sam} Morgan develops a consistent extension of the Bogoliubov theory
to higher order that leads to a gapless excitation spectrum.  This theory treats
the cubic and quartic terms of the Hamiltonian using perturbation theory in the
quasiparticle  basis. Expressions for the energy-shifts of the
excitations are given in Sec.~6.2 of Ref.~\cite{sam}, and we have calculated these shifts
for our simulations, with typical results plotted in Fig.~\ref{fig:full_theory}.

The energy spectrum predicted by the second order theory for the $\tilde{E} = 4000
$ simulation is in good agreement with the quasiparticle populations extracted from
the simulations, and is a significantly better fit than the Bogoliubov theory of
Eq.~(\ref{eqn:bog}).  
The validity of the second order theory 
is constrained by the requirement \cite{sam}
\begin{equation}
\left(\frac{k_B T}{n_0 U_0}\right)(n_0 a^3)^{1/2} \ll 1,
\label{eqn:validity}
\end{equation}
where $n_0$ is the condensate density.  For the results of
Fig.~\ref{fig:full_theory} with $\tilde{E} = 4000$, this parameter is 0.14 and
so we are beginning to probe the boundary of validity of the theory.
At higher $\tilde{E}$, the shifts it predicts at low $k$ are of the
    order of the unperturbed
energies, and the results become unreliable. In this region higher order terms
are important and the second order theory can no longer be expected to give
good results.

In summary for the system with $C_{\rm} nl = 2000$, Bogoliubov theory gives a
good prediction of the energy spectrum for simulations with total energies
$\tilde{E} \le 1600$, while the predictions of second order theory are good up
until about $\tilde{E} \approx 4000$.  We would like to point out, however,  that as
the GPE is non-perturbative we expect it will be valid up to and beyond the
transition region as long as the condition $N_k \gg 1$ is satisfied.

%Another quantity of interest is the vorticity of the system in equilibrium.  It
%has been argued that vortices may be important in the superfluid transition of
%$^4$He, reducing the superfluid density near the transition point
%\cite{Williams}.  We find that vortex rings do not appear in the simulations
%until $\tilde{E} > 7000$---precisely the region where Fig.~\ref{fig:n0ve}
%departs from linearity.  The same curve for the ideal gas ($C_{\rm nl} = 0$),
%however, is linear up to the critical region, suggesting that vortices may
%indeed play a role in the transition. In addition, it appears that the failure
%of the second order theory is linked to the appearance of vortices. We shall
%address these issues in both two and three dimensions in a future paper. 

In addition to the results described above, we have also run simulations with
$C_{\rm nl} = 10000$ and carried out an identical analysis.  We have found that
the results from evolving the GPE are qualitatively the same, and for very large
condensate fractions Bogoliubov theory accurately predicts the energy spectrum
accurately.   However, it appears that the second order theory develops a gap in
the energy spectrum in systems with a momentum cutoff.  This feature is yet to be
understood.

 In this Letter we have presented results for some of the equilibrium properties
of the homogeneous gas.   Other properties such as fluctuations, coherence
lengths, as well as the non-equilibrium dynamics  will be considered elsewhere. 
We would like to emphasise that this method relies on the lowest energy modes of
the system being classical in nature, and thus cannot handle situations where
strong quantum fluctuations are important.

In conclusion, we have presented evidence that the projected \GPE is a good
approximation to the classical modes of a Bose gas.  We have described how to
carry out the projection technique in the homogeneous case with periodic
boundary conditions, and have shown that starting with a randomised wave
function with a given energy, the projected GPE evolves towards an equilibrium
state.  We have analysed the numerical data in terms of the gapless, finite
temperature theory of Ref.~\cite{sam} in the classical limit, and found that
both the occupation and energies of the quasiparticles agree quantitatively 
with the predictions.

We would like to thank R.~J.~Ballagh and C.~W.~Gardiner for useful discussions.
This work was financially supported by St John's College and Trinity College,
Oxford, and the UK-EPSRC.

\end{multicols}
\end{document}